# Deep Learning for Longitudinal Medical Imaging: A Scoping Review

Author Names: Francesca Mussa[1,2+]; Divyanshu Tak[1,2+]; Atlas H. Avval[4], Sarah Brueningk[5,6], Ray H. Mak[1,2,3], Hugo J. W. L. Aerts [1,2,3], Andreas M Rauschecker[4], and Benjamin H. Kann[1,2,3]*

Author Affiliations:

1. Artificial Intelligence in Medicine (AIM) Program, Mass General Brigham, Harvard Medical School, Boston, MA, United States
2. Department of Radiation Oncology, Dana-Farber Cancer Institute and Brigham and Women's Hospital, Harvard Medical School, Boston, MA, United States
3. Dana-Farber Cancer Institute, Boston, MA, United States
4. Center for Intelligent Imaging, Department of Radiology & Biomedical Imaging University of California, San Francisco (UCSF)
5. Department of Department of Radiation Oncology, Inselspital, Bern University Hospital and University of Bern, Bern, Switzerland
6. Department of Digital Medicine, University of Bern, Bern, Switzerland

+ First Author
* Corresponding author
Correspondence address to:
Benjamin H. Kann, M.D.
Artificial Intelligence in Medicine (AIM) Program,
Mass General Brigham,
Harvard Medical School, 221 Longwood Avenue, Ste 442, Boston, MA 02115, MA, USA
Email: Bkann@bwh.harvard.edu

**List of abbreviations:**

AD – Alzheimer's disease

AI – Artificial Intelligence

AMD – Age-related macular degeneration

C – Classification

CFP – Color Fundus Photography

CNN – Convolutional neural networks

CT – Computed tomography

DR – Diabetic retinopathy

GRUs – Gated recurrent units

LLM – Large Language Model

LSTMs – Long short-term memory units

MRI – Magnetic resonance imaging

MS – Multiple sclerosis

OCT – Optical coherence tomography

ODP – Optic Disc Pit

R – Regression

RNN – Recurrent neural networks

SA – Survival Analysis

S – Segmentation

WCE – Wireless capsule endoscopy

## ABSTRACT

Longitudinal medical imaging analysis is a cornerstone of modern medical practice and patient care. Deep learning applied to longitudinal imaging offers wide potential to enhance diagnosis and track disease progression by capturing spatial changes over time. With major advances in single-timepoint deep learning for imaging, there has been growing interest in longitudinal image analysis, given its increased clinical relevance, though technical challenges remain. Several recent innovations may lead to a new era of multi-timepoint image evaluation, yet the scientific landscape, recent progress, and areas of need remain under-characterized.

To address this gap, we conducted a scoping review of deep learning methodologies applied to longitudinal medical imaging, yielding 102 studies published between 2018 and 2025. Neurological disorders (48%) and ophthalmic conditions (12%) were the most common clinical applications, with MRI serving as the predominant imaging modality (67%). Sequential feature modeling approaches combining convolutional neural networks (CNNs) with temporal models (LSTM/RNN) were the most frequent methodology (40%), followed by direct feature aggregation across timepoints (23%). Most studies targeted classification tasks (56%), while external validation was performed in only 24% of studies. Our findings highlight that deep learning-based longitudinal imaging analysis remains a promising field, though newer temporal architectures and larger datasets may improve success and clinical adoption of these tools.

**Trial registration:** OSF Registries, DOI: 10.17605/OSF.IO/5MGFE. Registered on Feb 18, 2026.

## INTRODUCTION

The integration of artificial intelligence (AI) into medical imaging has introduced a new era of possibilities in healthcare. Medical imaging techniques such as computed tomography (CT), magnetic resonance imaging (MRI), optical coherence tomography (OCT), and histopathology play vital roles across diseases, providing physicians with rich visual information for diagnosis and patient management.[1] Each imaging modality serves a distinct purpose: for instance, CT scans offer detailed cross-sectional views of internal structures, useful for pre-surgical planning or detecting intra-abdominal abnormalities, while MRI provides excellent soft-tissue contrast, ideal for assessing neurological and musculoskeletal disorders.[2] In clinical practice, physicians often compare current scans with prior examinations to evaluate disease progression, treatment response, or recurrence, relying on the patient's imaging history to contextualize findings.[3] However, as imaging archives expand, comprehensively assessing subtle longitudinal changes across multiple timepoints becomes increasingly difficult. Therefore, developing methods tailored to extraction and interpretation of spatiotemporal information from longitudinal imaging is critical to advancing clinical decision-making.

Over the past decade, deep learning has transformed medical image analysis, demonstrating remarkable success across a wide range of diagnostic and prognostic tasks.[4] These advances have largely centered on single timepoint data, detecting lesions, classifying mutations, and predicting outcomes from a single scan, showcasing the power of AI and data-driven approaches to extract complex imaging biomarkers that were previously inaccessible through manual analysis.[32] Yet, single timepoint evaluation is inherently incomplete, as it misses clinical context and overlooks the temporal dynamics of disease that are routinely leveraged by clinicians in real-world practice[5,6].

Longitudinal medical imaging, by incorporating spatial and temporal information over time, provides a richer foundation for capturing disease trajectories, identifying subtle progression patterns, and enabling precision diagnostics and prognostics.[5,32] However, applying deep learning to longitudinal imaging presents unique challenges such as variability in acquisition intervals, heterogeneous imaging protocols, and the need for frameworks capable of integrating spatiotemporal context without overfitting in data-limited settings[7,8]. Moreover, the current literature remains fragmented, often limited to small cohorts or single institutions with limited external validation, hindering generalizability and clinical translation [9,10,11].

Existing longitudinal learning frameworks include sequential feature modeling with recurrent neural networks or transformers, direct feature aggregation across time points, and generative approaches for disease trajectory prediction [9,12]. Yet, while promising, the extent to which these approaches have been validated, compared, or deployed in different real-world settings remains unclear.

To address this knowledge gap, we conducted a systematic scoping review of deep learning applications for longitudinal medical imaging across multiple clinical domains and imaging modalities. We characterize methodological trends, validation practices, and identify gaps that must be addressed to enable clinical adoption. By consolidating current evidence, we aim to characterize the current state of longitudinal deep learning in medical imaging, identifying key areas of progress and unmet needs in the field.

# RESULTS

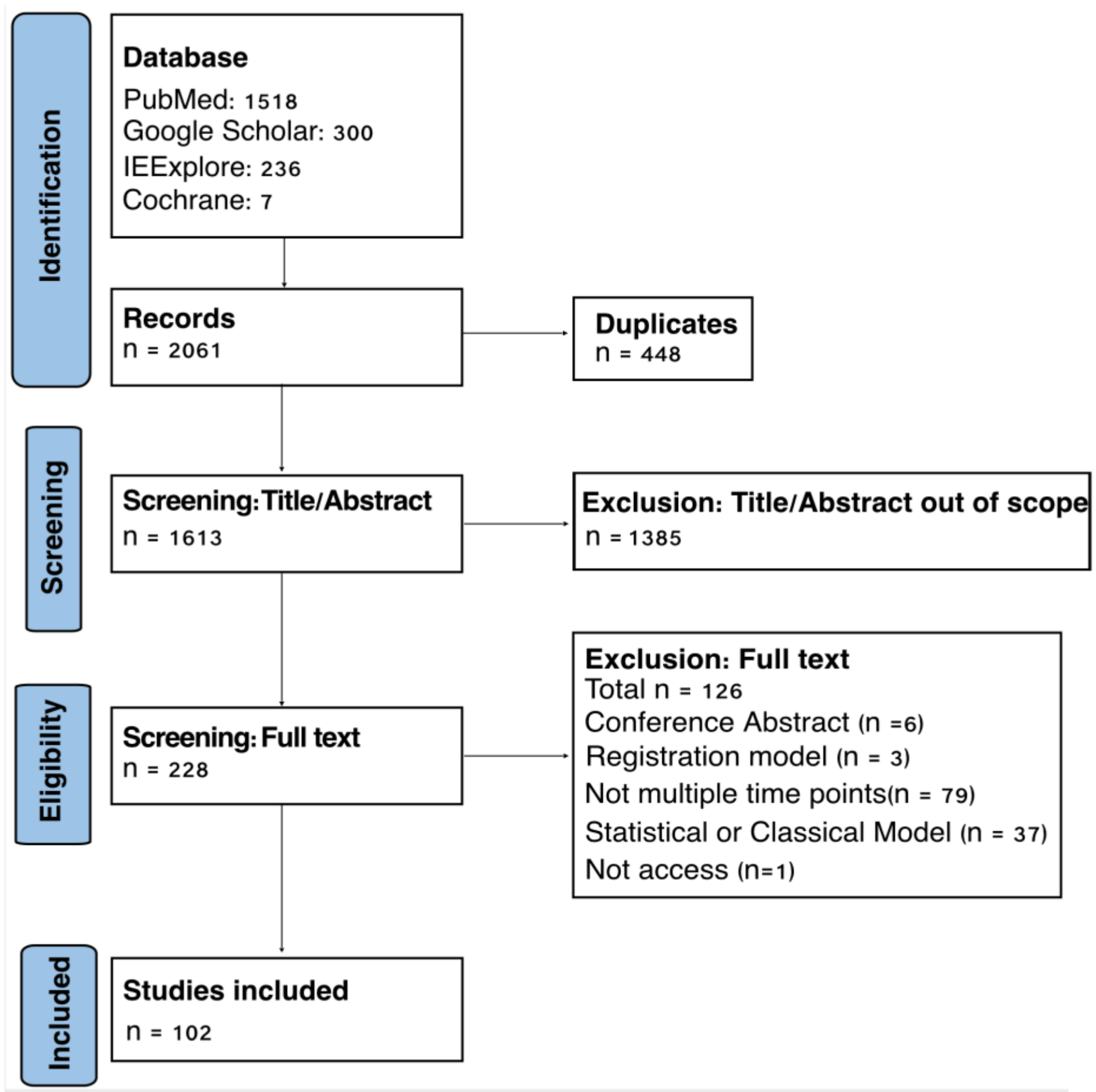

**Figure 1** PRISMA flow diagram illustrating the study selection process for this scoping review.

## Characteristics of studies

Following the screening of 1744 records, 228 full-text articles were assessed for eligibility (see "Methods"). The predominant reason for exclusion was the use of models trained on single time-point images, with outputs subsequently compared as if representing different disease stages, which did not meet the criteria for true longitudinal analysis. Ultimately, 102 studies met all inclusion criteria and were included in the scoping review (Figure 1).

Table 1 summarizes the characteristics of the studies included. Most of the studies were conducted by research groups based in the USA (41%), followed by China (32%) and Germany (9%). The datasets predominantly originated from North America (Figure 2d). All articles were published within the past eight years, with a peak in publications occurring in 2024 (Figure 3d). Regardless of whether the dataset was collected by the research team or obtained from open-source repositories, the sample sizes typically ranged from 30 to 3,000 participants.

Deep learning applied to sequences of medical images was predominantly used to investigate conditions affecting seniors (65+) (Figure 3b). Longitudinal medical imaging studies commonly leveraged publicly available datasets (57.5%). Regardless of dataset type, external validation of algorithms remained a limitation, with 76.5% studies lacking external validation (Figure 2c). 3% of included studies used a combination of public and private datasets to achieve a sample size for model training or to solve the lack of external validation data.

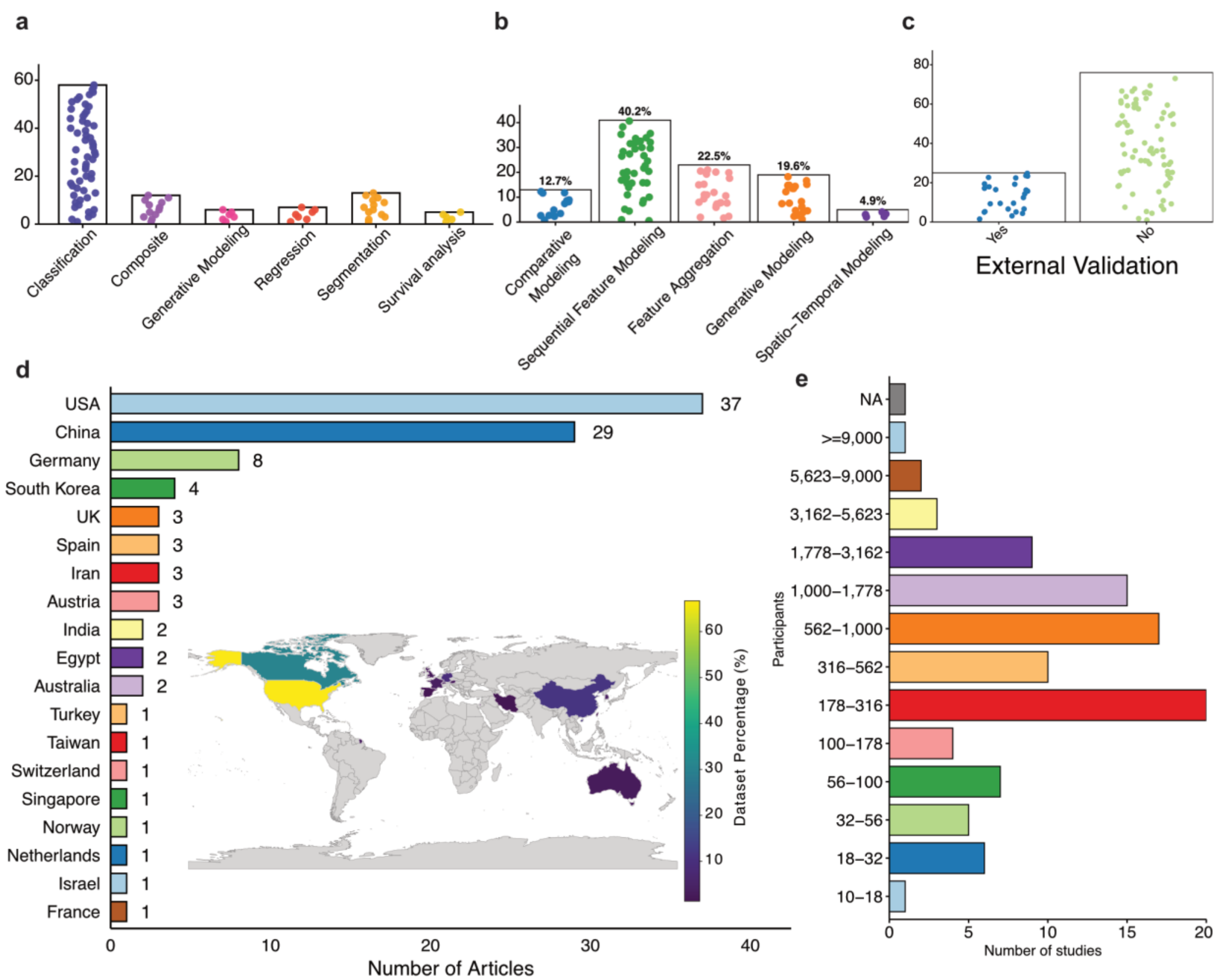

**Figure 2:** Study characteristics across aggregated longitudinal deep learning studies. (a) Distribution of study

outcome objectives, including classification, composite outcomes, generative modeling, regression, segmentation, and survival analysis. (b) Distribution of deep learning frameworks, encompassing feature aggregation, spatio-temporal modeling, pairwise comparative modeling, sequential feature modeling, and generative modeling. (c)

Distribution of studies with external validation. (d) Geographic distribution of studies by country of the leading research group, inset depicts the dataset origin distribution across countries. (e) Distribution of studies according to the number of participants per dataset, grouped into custom-defined bins.

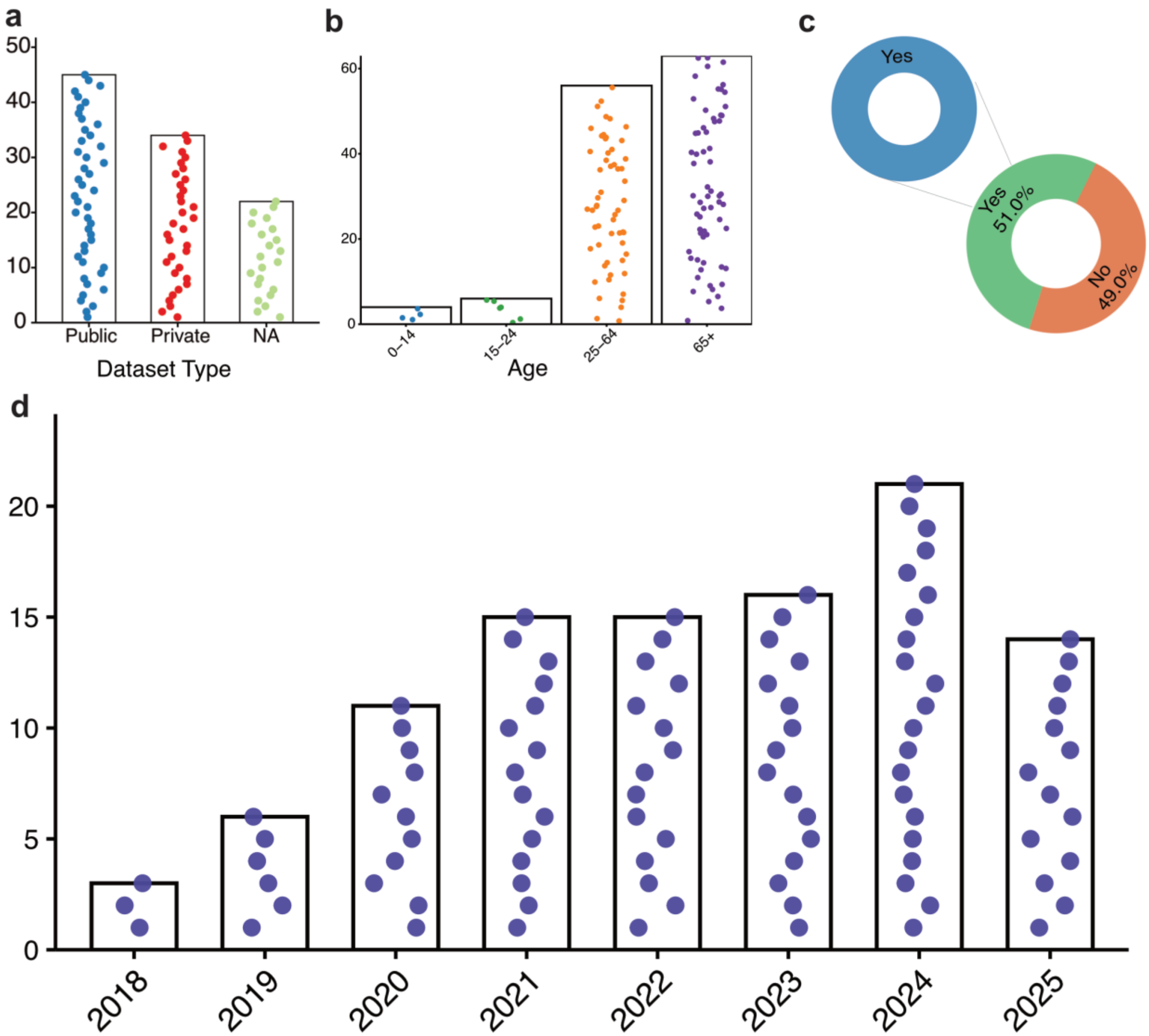

**Figure 3:** Dataset and study characteristics across aggregated studies. (a) Distribution of studies based on dataset accessibility, categorized as public, private, or not reported. (b) Distribution of study populations across age categories (0–14, 15–24, 25–64, and ≥65 years), indicating predominant focus on adult and older adult cohorts. (c) performance comparisons between longitudinal and single-timepoint model. The outset shows the proportion of studies that conducted direct comparisons (48.3%), while the inset shows that longitudinal approaches outperformed single-timepoint baselines in all of these studies. (d) Distribution of publications across years (2018-2025)

## Brain and ophthalmic imaging predominate longitudinal deep learning

Deep learning has been applied to a sequence of medical imaging across multiple anatomical regions. In 2018, only three studies were published, one focusing on neurological disorders and the other on gastrointestinal conditions. Subsequently, approaches were extended to other body parts and medical conditions. The number of studies on neurological disorders peaked between 2021 and 2022, with the majority focusing on Alzheimer’s disease (AD)(78%). From 2022 to 2025, 41% of the studies across medical condition categories applied deep learning to cancer-related data (Figure 4b).

Across conditions and anatomical sites, 78% of studies used sequences of two to four scans to train deep learning models for tasks such as classification and composite outcomes (Figure 4a). Segmentation tasks were mainly performed on brain-related conditions (54%), although some studies applied it to heart failure (8%) and breast/lung cancer datasets (23%). The required sequence length for segmentation varied depending on the specific medical condition.

Approximately 48% of the included studies focused on brain-related conditions, particularly Alzheimer’s Disease (61%) (Figure 5). Within these studies, most addressed classification tasks and typically included a maximum of 2000 participants. In neurological disorder research, sequence length for classification, regression, and generative modeling tasks tended to increase with larger sample sizes, while studies combining classification and regression usually required sequences of more than three scans. Generative modeling was exclusively applied to neurological disorders (65%), such as multiple sclerosis and Alzheimer's, potentially due to its effectiveness in monitoring disease progression (Figure 4a).

Ophthalmic conditions represented the second most studied field (12%), closely followed by lung conditions (11%), with age-related macular degeneration (AMD) accounting for 50% of the included eye-related studies. (Figure 5) Research in AMD peaked between 2019 and 2021, after which the focus shifted primarily toward glaucoma. (Figure 4b) In contrast to neurological disorders, studies on eye conditions addressed a narrower range of tasks, with one study examining time-to-event analysis, one focused on regression modeling and another incorporating regression into classification. For classification tasks, sequence lengths differed across studies. (Figure 4a) For example, Mohammadzadeh *et al.* (2023) used a minimum of six longitudinal scans for glaucoma classification using optical disc photographs (ODP), whereas Rivail *et al.* (2019) employed two scans for AMD classification using optical coherence tomography (OCT). (Table 1)

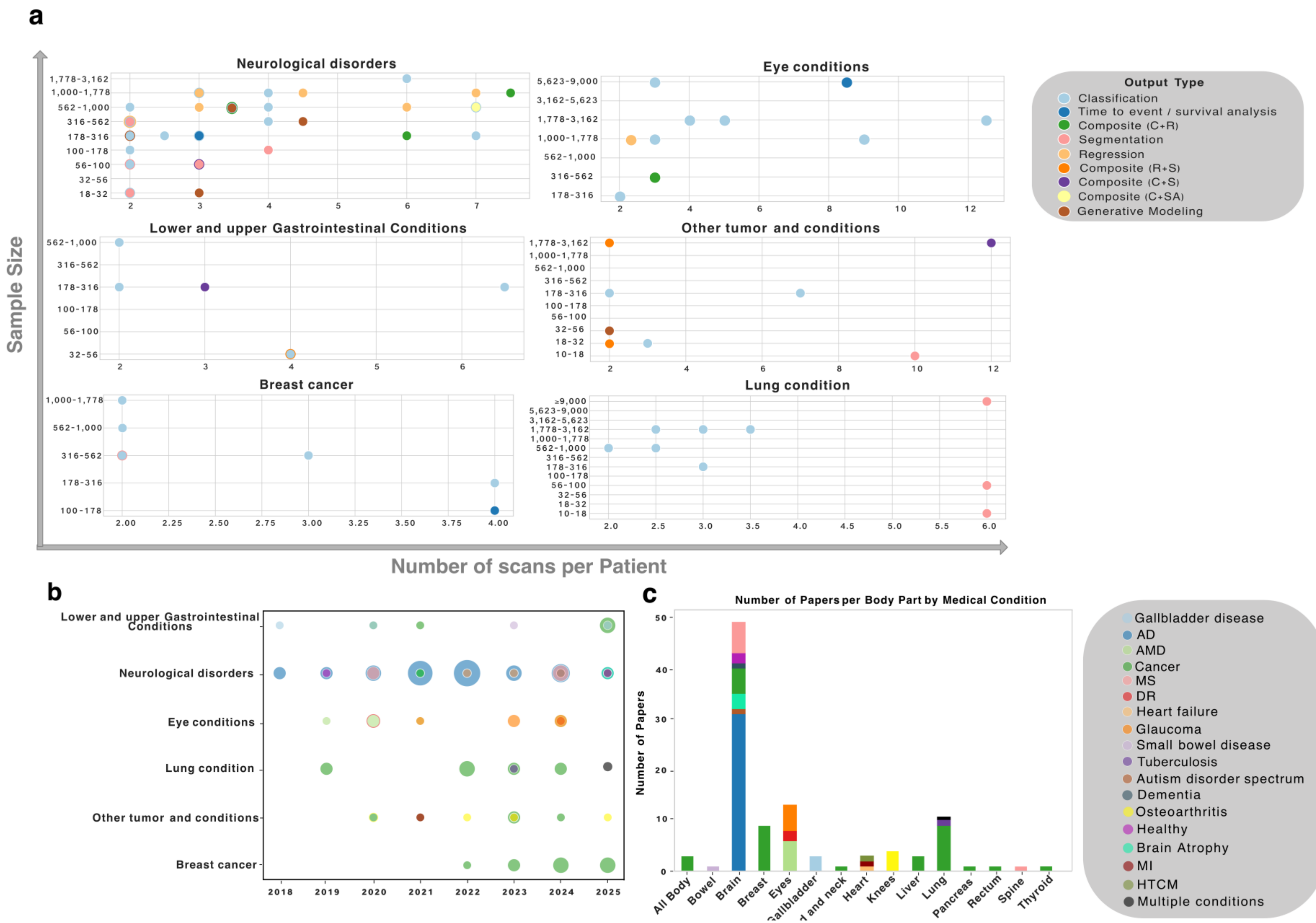
a
Sample Size
Number of scans per Patient
Neurological disorders
Eye conditions
Lower and upper Gastrointestinal Conditions
Other tumor and conditions
Breast cancer
Lung condition
Output Type
Classification
Time to event / survival analysis
Composite (C+R)
Segmentation
Regression
Composite (R+S)
Composite (C+S)
Composite (C+SA)
Generative Modeling
b
Lower and upper Gastrointestinal Conditions
Neurological disorders
Eye conditions
Lung condition
Other tumor and conditions
Breast cancer
2018
2019
2020
2021
2022
2023
2024
2025
c
Number of Papers per Body Part by Medical Condition
Number of Papers
All Body
Bowel
Brain
Breast
Eyes
Gallbladder
Head and neck
Heart
Knees
Liver
Lung
Pancreas
Rectum
Spine
Thyroid
Gallbladder disease
AD
AMD
Cancer
MS
DR
Heart failure
Glaucoma
Small bowel disease
Tuberculosis
Autism disorder spectrum
Dementia
Osteoarthritis
Healthy
Brain Atrophy
MI
HTCM
Multiple conditions

**Figure 4. Medical Conditions.** Scatter plots (a) show the relationship between sample size and number of scans per medical condition, highlighting output type (C+R= Classification + Regression, C+R= Classification + Regression, R+S = Regression + Segmentation and C+SA= Classification + Survival Analysis). The bubble plot (b) illustrates the number of studies published over the last eight years, stratified by medical condition. The bar chart (c) shows the number of papers focusing on specific medical conditions (AD = Alzheimer's disease, AMD = age-related macular degeneration, MS = multiple sclerosis, DR = diabetic retinopathy, MI = Myocardial infarction, HTCM = Hypertensive cardiomyopathy) for each body part, with "Multiple Conditions" referring to studies whose datasets include more than one distinct medical condition.

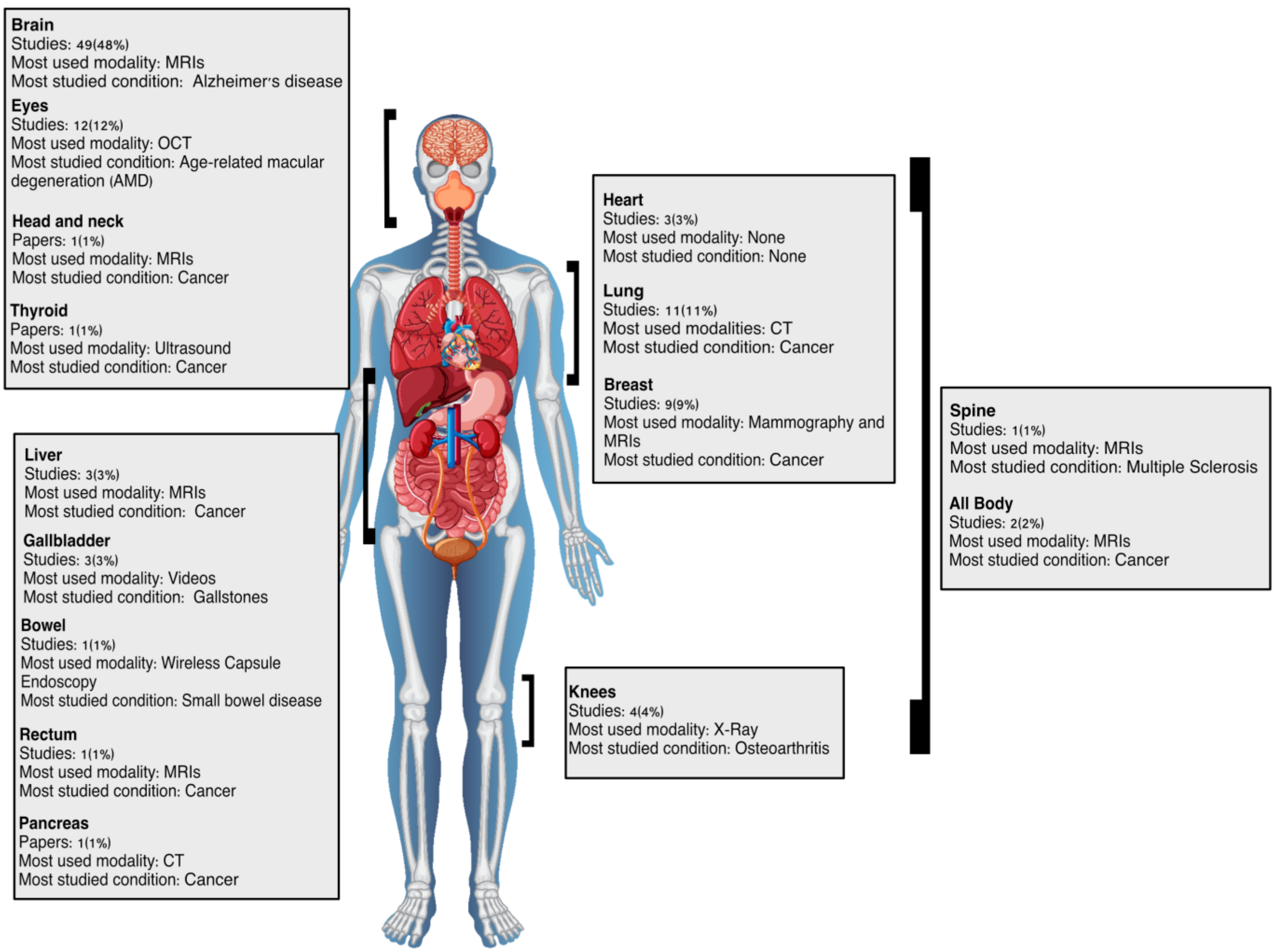

**Figure 5. Overview of studies associated with different body parts.** Whole-body image divided by body parts, highlighting the number of studies, the most frequently addressed conditions, and the most used modalities in the included studies. "None" indicates that no single condition or imaging modality predominated across the studies.

## Emphasis on MRI

The most frequently used imaging modality was MRI, accounting for approximately 67% of the studies included, followed by CT (10%) (Table 1). There was a notable gap in available open-source datasets across different imaging modalities, with MRI having the largest number of accessible datasets such as Alzheimer's Disease Neuroimaging Initiative (ADNI) and the Open Access Series of Imaging Studies (OASIS).(Fig.4a) In contrast, all imaging modalities for eye-related conditions lacked publicly available longitudinal datasets, despite considerable research interest in areas such as age-related macular degeneration (AMD) using longitudinal deep learning approaches. This scarcity limits external validation of deep learning approaches.

Only three of the included studies used different imaging modalities for the same medical condition, specifically fMRI combined with MRI and MRI combined with PET, both used for generative modeling tasks, and one study combining MRI, PET, and DWI for sequential feature modeling. In contrast, only one study (Li et al., 2020) applied the same deep learning model to two different conditions, using both Color Fundus Photography (CFP) and X-ray images (Table 1). Across all included studies, MRI was the only imaging modality used for all types of tasks (Figure 6c), with the length of sequences varying according to the specific task. Studies employing transformer-based generative models typically used short longitudinal sequences, most commonly three to four scans across medical imaging modalities. (Figure 6b) Overall, MRI remained the preferred modality that combined multiple tasks, such as classification and regression in terms of predicting progression and outcome changes in oncologic and neurocognitive applications. (Figure 6c)

In terms of model architecture, direct feature aggregation and sequential feature modeling were the most frequently used architectures in the included studies using MRI data, accounting for 28% and 40%, respectively. (Table 1) Approximately 52% of MRI-based studies applied these architectures to investigate Alzheimer's disease. In contrast, transformer-based and generative modeling approaches using MRI were also applied to datasets of healthy participants, primarily to study brain connectivity. (Table 1)

In studies of ophthalmic conditions, the ODP datasets were employed for glaucoma classification. However, despite addressing the same task and using the same model type, Mohammadzadeh et al. (2024) used three time points, whereas Vahid Mohammadzadeh et al. (2023) used an average of 12 scans per patient. In contrast, Color Fundus Photography (CFP) and Optical Coherence Tomography (OCT) modalities were predominantly utilized for feature extraction and generative modeling applications. (Table 1)

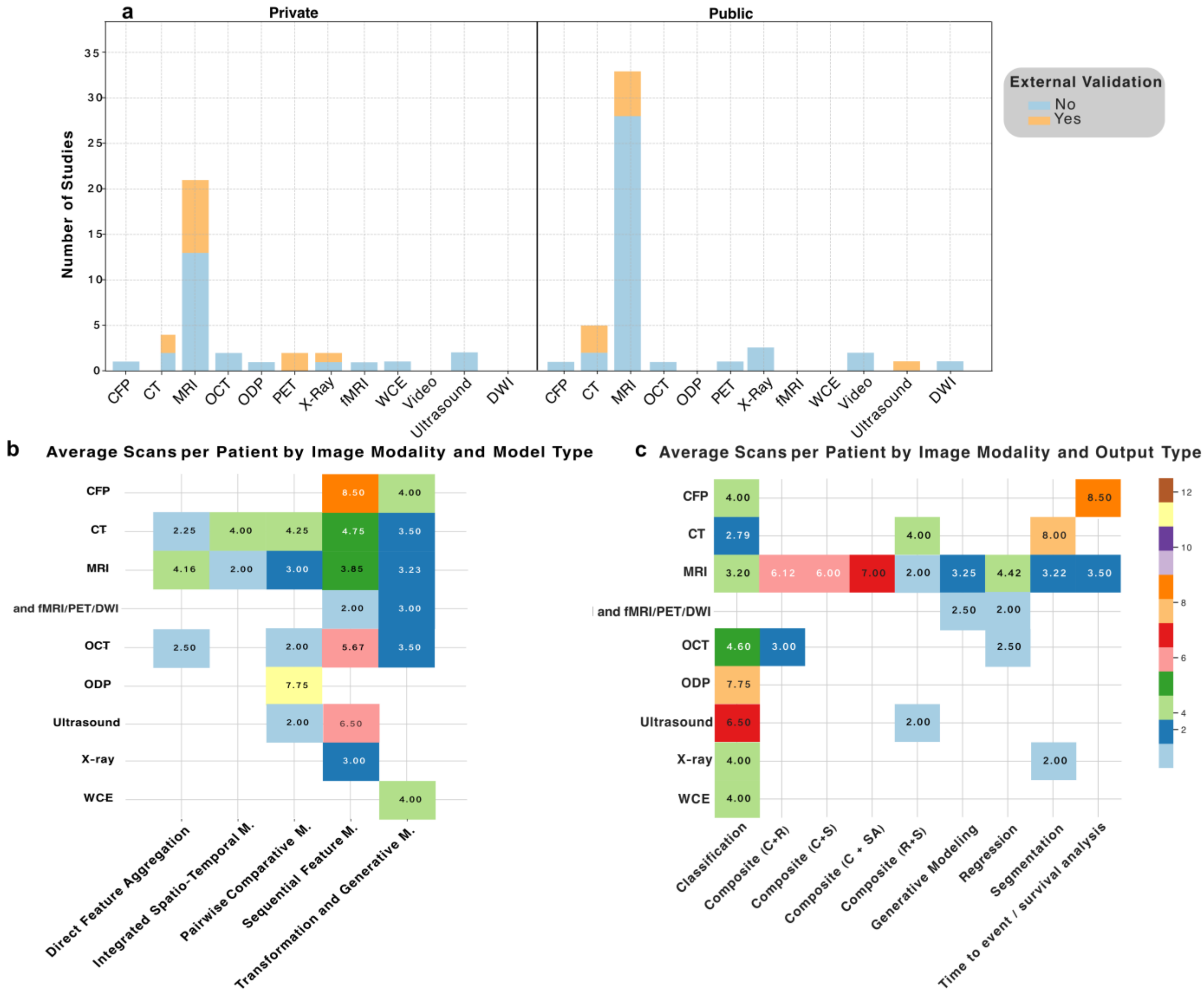

**Figure 6. Imaging Modalities.** (a) Bar chart illustrating the number of studies employing each medical imaging modality, stratified by dataset type (public, private, or combined) (b) Heat map depicting the average sequence length by imaging modality and model architecture (C+R= Classification + Regression, C+R= Classification + Regression, R+S = Regression + Segmentation and C+SA= Classification + Survival Analysis). (c) Heat map illustrating the average sequence length by imaging modality and output type.

## Classification was the most common use case

Classification (~56%) constituted the most common primary task across the reviewed literature (Figure 2a), representing the dominant focus of longitudinal deep learning research. Segmentation, regression, survival analysis, and composite modeling objectives were comparatively uncommon. This outcome distribution underscores the field's current emphasis on categorical diagnostic and disease-state classification while revealing a critical gap in the development of methods for prognostic modeling and longitudinal prediction of disease trajectories. Among the studies reviewed, around half (48%) directly compared longitudinal models to single-timepoint baselines and demonstrated improved performance with longitudinal approaches (Figure 3c). This indicates that incorporating temporal information can enhance predictive performance, although the benefit remains dependent on cohort size, follow-up duration, and temporal consistency across scans.

### CNN-RNN combinations are the most used architectures

Sequential feature modeling is employed as the predominant framework across longitudinal deep learning studies (Figure 2b), applied in the largest proportion of works (~40%). This approach integrates convolutional neural networks (CNNs) for spatial feature extraction with temporal architectures such as recurrent neural networks (RNNs), long short-term memory (LSTM) units, gated recurrent units (GRUs), or transformer- encoders, to capture change in features across time points. The sample sizes in these studies varied widely but were generally concentrated in the mid-range cohorts (several hundred to a few thousand participants; Figure 2e, Figure3a,b), demonstrating the high data requirements for training in this approach. Notably, sequential feature modeling has shown a steady increase over time, while direct feature aggregation also rose but stabilized from 2023 onwards. By contrast, generative modeling approaches, designed to synthesize subsequent time-point images, were less common but displayed applications of adversarial networks and autoencoders, often requiring larger, high-dimensional datasets to achieve stable training, with a declining trend observed after 2023. Feature aggregation strategies, which merge spatial features from vision encoders into a unified representation using non-learning operations (e.g., mean pooling, max pooling, addition), were employed in a moderate number of studies, particularly when datasets were smaller and longitudinal depth was limited. Pairwise comparative frameworks, built on Siamese networks to quantify similarity or progression between two scans, were applied in smaller-scale cohorts (frequently <1,000 participants), demonstrating efficiency in scenarios with sparse longitudinal data. Finally, spatio-temporal modeling approaches, which jointly encode spatial and temporal dynamics, were predominantly used in 2020 and 2021 but were largely absent in subsequent years, and were rarely employed overall. Across studies, architecture selection appeared to reflect the structure of the longitudinal task rather than disease type alone: sequential models were most common for ordered prediction tasks, feature aggregation for limited-scan settings, pairwise models for two-timepoint change detection, and generative or spatiotemporal models for progression modeling, synthesis, or segmentation.

## DISCUSSION

Deep learning for longitudinal imaging represents a small, but growing focus of computational biomedical imaging research over recent years[8]. Our scoping review identified 102 studies focusing on longitudinal imaging, most of which were conducted within the past few years. Compared to the single timepoint deep learning for medical imaging studies published, this number is exceedingly small. Our scoping review of those 102 studies published between 2018 and 2025 reveals four key findings. First, neurological disorders, particularly Alzheimer's disease, dominate this field, reflecting both clinical interest and the accessibility of well-curated longitudinal neuroimaging datasets. Notably, a small subset of studies (~7.8%, such as Aslani et al., 2024 and Du et al., 2025) utilized public benchmarks as development sandboxes followed by validation on independent private cohorts. Research volume is strongly correlated with public data availability, as 60% of neurological studies utilize open-source repositories like ADNI. Conversely, the lack of standardized public longitudinal data in fields such as ophthalmology (where only 15% of studies were public) remains a significant barrier to benchmarking and external validation (Supplementary Table S1). Second, sequential feature modeling, by combining convolutional neural networks with temporal architectures such as LSTMs or transformers, represents the predominant methodological paradigm by capturing the temporal dynamics in disease characterization. Third, the majority of studies focus on diagnostic classification rather than prognostic or trajectory-based modeling, highlighting a research gap in predicting future disease evolution and prognostication. Fourth, external validation remains very limited, only 24% of studies conducted multi-institutional or cross-dataset evaluation, making it a fundamental barrier to clinical translation and generalizability. Our findings highlight that perhaps the most important barrier to development of longitudinal models is a lack of longitudinal training and validation datasets[9,10,11].

Within specific domains, neurological imaging, especially Alzheimer's disease and multiple sclerosis has served as the primary applications for longitudinal deep learning [16,46,56]. Notably, studies demonstrate that sequences as short as two to three MRI scans can yield meaningful performance for diagnostic tasks, while larger cohorts and longer sequences enhance sensitivity to subtle temporal changes. In ophthalmology, age-related macular degeneration and diabetic retinopathy have emerged as key applications, with generative models showing promise in synthesizing follow-up retinal images to monitor disease progression[45,58]. However, beyond these select areas, applications remain sparse, with limited work in oncology, cardiology, or pediatrics where longitudinal imaging could be particularly impactful for early response assessment or developmental trajectory modeling. Among studies that directly compared longitudinal models with single-timepoint baselines, longitudinal approaches generally showed numerically improved performance according to each study's task-specific metric; however, the magnitude of improvement and statistical significance were inconsistently reported, limiting direct cross-study comparison.  Only six studies leveraged multimodal longitudinal learning, published between approximately 2021 and 2025, combining imaging with associated clinical data. Such approaches integrated MRI with demographic variables (e.g., age, sex, education), structured clinical features and pathology, or richer clinical data streams such as vital signs, laboratory tests, medications, ophthalmic examinations, optical coherence tomography derived features, and clinical notes. Only one study combined chest X-ray with clinical data for automated radiology report generation. These studies demonstrated improved predictive performance over imaging-only models [22,33,62] .There is a clear need for research in medical conditions outside of brain and eye disease. There may be particular opportunities for longitudinal analyses for diseases wherein longitudinal screening is utilized, such as breast cancer and lung cancer.

Methodologically, the field is dominated by sequential and feature aggregation frameworks, while spatiotemporal modeling and generative paradigms remain underexplored despite their potential for capturing nonlinear disease dynamics. Most existing models rely on modest sample sizes (typically 30–3,000 patients) and homogeneous datasets, which limits generalization across institutions, scanners, and acquisition protocols. The underrepresentation of pediatric and multimodal imaging cohorts further limits the generalizability of across wide ranging patient population. There was a lack of investigation of vision transformers and vision language models, which was surprising, given their popularity in the broader deep learning field and there suitability for sequence-based data. Technical challenges and ideal datasets for validation may have limited study thus far. Notably, the temporal evolution of architectural choices mirrors broader trends in the deep learning community, progressing from simple feature aggregation strategies and recurrent models such as LSTMs and RNNs, toward transformer and attention-based architectures, paralleled by the recent emergence of diffusion-based generative models. Overall, a clear evolutionary trajectory can be framed as follows: from simple pooling and concatenation, to recurrent models such as LSTMs and RNNs, and more recently toward transformer and attention-based architectures. Ultimately, the field would benefit from the establishment of standardized longitudinal benchmarks and public challenges, analogous to the BraTS challenge for segmentation, to catalyze research and enable objective methodological comparisons across diverse disease trajectories.

This review benefits from a systematic and reproducible process guided by the scoping review framework. In addition, the screening of studies was conducted by two independent reviewers. However, this study has limitations. The search strategy focused only on publicly available articles, and literature such as preprints and abstracts was not included. As a result, certain potential problems or challenges in applying deep learning to longitudinal medical imaging may not have been captured.

Looking forward, longitudinal deep learning in medical imaging is likely to move from task-specific models trained on small homogeneous cohorts toward larger, externally validated, and benchmarked frameworks that can model disease trajectories across institutions. Progress will likely depend on standardized longitudinal datasets and open challenges, stronger temporal architectures such as transformers and foundation models, and multimodal approaches that integrate imaging with clinical context. These developments may shift the field from retrospective classification toward clinically actionable prediction of progression, treatment response, and future disease states.

To conclude, this study represents the first comprehensive review of deep learning methodologies for longitudinal medical imaging, demonstrating that while significant progress has been made, the field remains in an early developmental stage marked by limited validation and narrow clinical focus. These findings summarize the state-of-the-art in this growing area of biomedical AI and form a foundation for future research aimed at building robust, generalizable, and clinically meaningful longitudinal models. By advancing temporal modeling in medical imaging, the field moves closer to developing AI systems capable of capturing disease evolution, guiding treatment decisions, and ultimately improving patient outcomes.

# METHODS

We conducted a scoping review following the framework proposed by Arksey and O'Malley, and in accordance with the PRISMA-ScR (Preferred Reporting Items for Systematic Reviews and Meta-Analyses extension for Scoping Reviews) guidelines [13], with reporting adherence documented in the accompanying PRISMA-ScR checklist (Supplementary Table S2). The review was prospectively registered on the Open Science Framework (OSF) (DOI: 10.17605/OSF.IO/5MGFE) on February 18, 2026. The following key questions are investigated in detail from the published studies:

(i) Which types of medical conditions are studied using deep learning in longitudinal medical imaging?
(ii) What are the most frequently used imaging modalities in longitudinal medical imaging?
(iii) Which deep learning approach is most used?
(iv) What are the most analyzed outcomes in longitudinal medical imaging?

Figure 1 shows the flow chart for study selection process.

### Search Strategy

A literature search was conducted across PubMed, Google scholar, IEEExplore and Cochrane library on January 22, 2026, covering studies published from 2018 to 2025. Two independent reviewers performed the search combining controlled vocabulary (MeSH terms) and free-text keywords related to deep learning, imaging modality, and longitudinal or temporal analysis. Modality-specific search strategies were developed for MRI, CT, radiography, ultrasonography, PET, endoscopy, optical coherence tomography, mammography, and electrocardiography. Given the absence of prior scoping reviews with a comparable scope covering vision-based deep learning models applied to longitudinal medical imaging across all modalities and body parts, the search strategy was independently designed to align with the specific objectives of this review. The complete search strategies are reported in Supplementary Data S1. To ensure that our search strategy was comprehensive, we verified that papers from major conferences (e.g., ISBI, BIBM, ICAIHI) appeared in the search results and were included.

### Study Selection

The duplicate records were first removed from the pool of articles aggregated across the 4 sources by the two reviewers. Post duplicate records removal, the remaining articles underwent a two-stage screening process by two independent reviewers, with predefined inclusion and exclusion criteria applied at both stages. In the first stage, titles and abstracts were screened for relevance to our study objective, specifically longitudinal deep learning applications in medical imaging. In the second stage, full-text articles were assessed based on predefined inclusion and exclusion criteria. To ensure methodological rigor and minimize bias, each reviewer independently screened the final set of full-text articles identified by the other reviewer. Any discrepancies between reviewers were resolved through discussion to reach consensus.

## Inclusion and Exclusion Criteria

The scoping review focused on studies that applied vision-based deep learning models to a sequence of at least two medical images from the same patient at different timepoints, irrespective of imaging modality, body part, or study objective. Inclusion and exclusion criteria were defined to prioritize approaches in which the primary model operated directly and exclusively on medical images. In line with these criteria, studies were excluded if they were not published in English, involved animal subjects, utilized large language models (LLMs), employed image registration–based models, were available only as preprints or conference/abstracts, or focused exclusively on pathology slides. Studies that relied solely on statistical or classical machine-learning models applied to pre-extracted features were also not considered. Only original research articles were eligible for inclusion, with systematic reviews, scoping reviews, and other secondary literature excluded. In addition, only published-peer reviews were included due to uncertainty regarding their final status.

## Extraction and aggregation of information

After screening, article- and study-level information was extracted and aggregated from the included studies (Table 1). Specific data were categorized according to predefined rationales to streamline analysis and synthesis (Table 2). Article-level characteristics included, the year of publication, and country of the research group. While the study-level characteristics included participant demographics, methodological approach, type of validation, study outcomes, and imaging data modalities. When information was unclear or unavailable, it was marked as indeterminate/missing (N/A) and excluded from the analysis. Data extraction was conducted using a standardized spreadsheet in Microsoft Excel. Extracted data were summarized using descriptive statistics, including frequencies and percentages, to characterize study distributions across key variables.

**Table 1.** Aggregated summary table of the 102 studies. NA represents that the corresponding information was not available in the publication. A study was classified as using a public dataset if at least one of the datasets employed was publicly available. Country of origin refers to the institutional affiliation of the first author. **(attached)**

| Variables | Aggregation Explanation |
|---|---|
| Age Class | Based on standard classifications: Children (0-14), Youth (15-24), Adults (25-64), and Seniors (65+)[14] |
| Imaging Modalities | All types of MRI are grouped together. Mammography was classified under X-ray based imaging due to shared acquisition technique, while echocardiography was classified under ultrasound. |
| Medical Condition | Diseases are grouped by affected body region and/or clinical similarity. 'Other tumor and condition' category includes diseases that do not fall into the primary classification groups |
| Deep Learning method | Grouped by presence of feature extraction method and whether the model captures temporal, spatial, or both relationships (see Supplementary Data S2) |

**Table 2.** Methodology for Aggregating Study-Level Characteristics.

## ACKNOWLEDGEMENTS

This study was supported in part by the National Institutes of Health (NIH) through grants U54 CA274516, R01 DE034780, and P50 CA165962 (B.H.K.). Additional support was provided by the Botha–Chan Low Grade Glioma Consortium and the DMG Precision Medicine Initiative (B.H.K.).

## AUTHOR CONTRIBUTIONS

Francesca Mussa[1,2+,]; Divyanshu Tak[1,2+,]; Atlas H. Avval[4,5], Sarah Brueningk[6,7], Ray H. Mak[1,2,3], Hugo J. W. L. Aerts [1,2,3], Andreas M Rauschecker[4,5], and Benjamin H. Kann[1,2,3]*

**F.M, D.T.** and **B.H.K.** designed the study, developed the methodology, and the screening criteria. **F.M** and **D.T.** performed paper screening and review along with the analysis, table creation and writing of the manuscript. **A.H.A**, **S.B**, **R.H.M**, **H.J.W.L.A**, and **A.M.R.** contributed to data interpretation critical discussion of the results. **B.H.K.**, **F.M** and **D.T.** revised the manuscript with input from all authors. **B.H.K.** supervised the study. All authors reviewed and approved the final manuscript.

## DECLARATION OF INTERESTS

The authors declare no competing financial or non-financial interests related to this work.

## DATA AVAILABILITY

All data generated or analyzed during this study is available from the corresponding author upon reasonable request.